\documentclass[runningheads]{llncs}

\usepackage[switch]{lineno}
\usepackage{xcolor}
\usepackage{cite}
\usepackage{colortbl}
\definecolor{lightergray}{gray}{0.9}
\usepackage{amsmath,amssymb,amsfonts}
\usepackage{algorithmic}
\usepackage[font=footnotesize]{caption}
\usepackage{subcaption}
\usepackage{graphicx}
\usepackage{textcomp}
\usepackage{xcolor}
\usepackage{hyperref}
\usepackage{xspace}
\usepackage{mathtools,tabularx}
\def\BibTeX{{\rm B\kern-.05em{\sc i\kern-.025em b}\kern-.08em
    T\kern-.1667em\lower.7ex\hbox{E}\kern-.125emX}}

\newcommand\tool{\textit{ARC-V}\xspace}

\begin{document}

\title{\textit{ARC-V}: Vertical Resource Adaptivity for HPC Workloads in Containerized Environments}
\titlerunning{Vertical Resource Adaptivity for HPC in Containerized Environments}
\author{Daniel Medeiros, Jeremy J. Williams, Jacob Wahlgren, Leonardo Saud Maia Leite, Ivy Peng}
\authorrunning{D. Medeiros, J. J. Williams, J. Wahlgren, L. S. M. Leite, I. Peng}

\maketitle

\begin{abstract}
Existing state-of-the-art vertical autoscalers for containerized environments are traditionally built for cloud applications, which might behave differently than HPC workloads with their dynamic resource consumption. In these environments, autoscalers may create an inefficient resource allocation. This work analyzes nine representative HPC applications with different memory consumption patterns. Our results identify the limitations and inefficiencies of the Kubernetes Vertical Pod Autoscaler (VPA) for enabling memory elastic execution of HPC applications. We propose, implement, and evaluate ARC-V. This policy leverages both in-flight resource updates of pods in Kubernetes and the knowledge of memory consumption patterns of HPC applications for achieving elastic memory resource provisioning at the node level. Our results show that ARC-V can effectively save memory while eliminating out-of-memory errors compared to the standard Kubernetes VPA.
\end{abstract}


\section{Introduction}
One of the distinctive features of traditional HPC systems is how infrastructure and computing resources are provisioned. It is common to provision entire bare-metal nodes within the local, on-premise system. One consequence of this approach is that the entire set of resources will be reserved for the user, even if not fully utilized, leading to potential resource waste, as complex HPC workloads often exhibit varying phases and different resource utilization patterns~\cite{peng2021holistic}. Meanwhile, in cloud-first environments, where containerized workloads are executed, it is usual to define the resources allocated to a specific workload in a finer granularity. This policy potentially allows multi-tenancy of workloads within the same node, while allowing the use of spare resources for scaling or running background processes.
\begin{figure*}
     \centering
     \includegraphics[width=\textwidth]{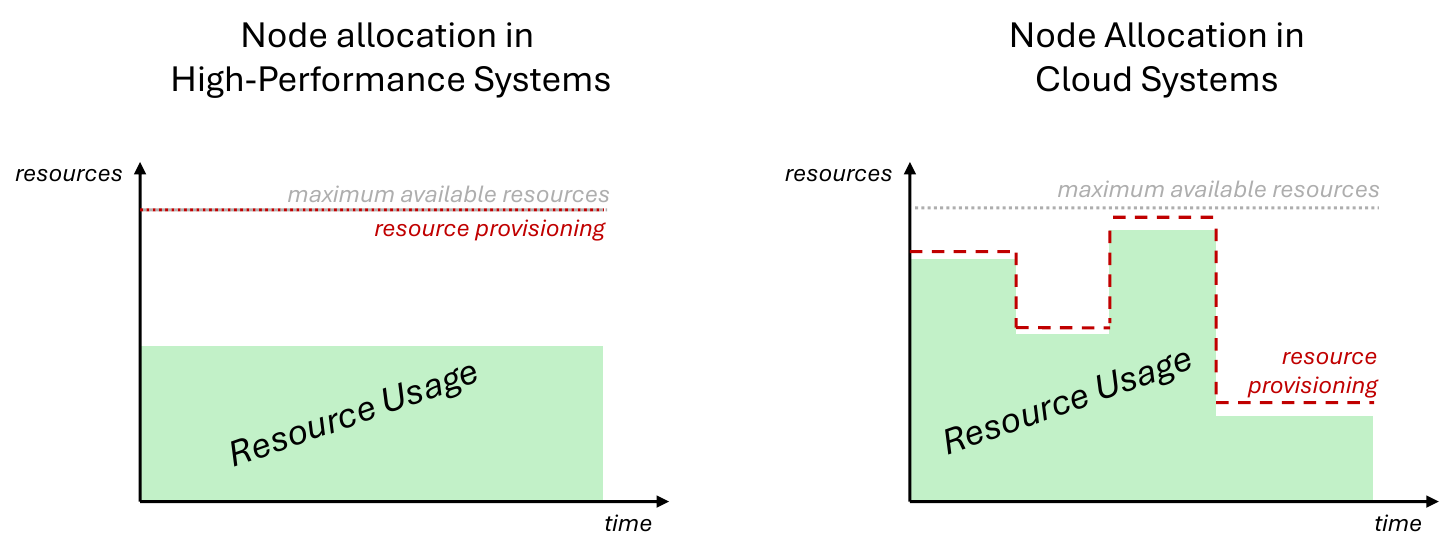}
        \caption{A high-level overview of how resource allocation for tasks in high-performance systems (HPC) and cloud systems generally works. In the former, allocations are static even if the resources are not fully used, while the latter displays a degree of flexibility for changing the allocations on execution time.}
        \label{fig:interesting}
\end{figure*}

The execution of HPC workloads in containerized environments requires adopting new strategies to fully leverage the benefits commonly associated with cloud workloads. A key distinction lies in application coupling: while cloud applications tend to be loosely coupled, HPC workloads are often tightly coupled. This tight coupling makes HPC applications highly sensitive to out-of-memory~(OOM) errors, as the default behavior of MPI-based applications means that a failure in a single node may cause the entire application to fail. Furthermore, until recently, changing allocated resources in a container during its execution meant that a container would have to be restarted, which creates a major problem for HPC workloads. Although checkpointing measures may mitigate such failures in certain cases~\cite{egwutuoha2012proactive, jin2012checkpointing}, these measures are not a universal solution and may lead to performance degradation in the workload.

A potential method for optimizing resource utilization for workloads in containerized environments is through the usage of a vertical autoscaler to manage the application's needs without reserving the entire node for it. However, many state-of-the-art autoscalers that were originally built for cloud workloads are not suitable for HPC workloads as different inputs may significantly alter the memory pattern, or they might not properly adapt to these workloads as HPC workloads might run for a shorter time than user-facing cloud services.


Therefore, this paper explores the feasibility of vertical resource adaptivity for the memory of containerized HPC workloads with dynamic memory consumption in a containerized environment. Our key contributions are:


\begin{itemize}
\item We discuss how HPC workloads may benefit from elastic memory scaling in cloud environments, and examine several patterns of memory consumption across nine HPC applications and benchmarks;
\item We identify the limitations of the state-of-the-art Kubernetes Vertical Pod Autoscaler in HPC workloads, and build a simulator for its scaling policy;
\item We propose, implement and evaluate the \textbf{A}daptive \textbf{R}esource \textbf{C}ontroller - \textbf{V}ertical (\tool) to achieve the resource adaptivity in memory through a reactive vertical autoscaler that does not need \textit{a priori} knowledge of the application, reducing the performance impact on HPC workloads in containerized environments in comparison to the Vertical Pod Autoscaler available in Kubernetes.
\end{itemize}


\section{Background \& Related Work} \label{sec:background}
\subsection{Kubernetes}

Kubernetes is a container orchestrator, responsible for coordinating how containerized workloads are distributed across the available nodes. Containers are encapsulated in an abstraction model called "pod," the smallest deployable unit within Kubernetes. A pod contains at least one container image and the resources (CPU, memory) that are allocated to it, and a pod may contain more than one container. Pods are designed with long-running workloads in mind and if they fail, Kubernetes, by default, attempts to recover them by restarting. 

Within a Kubernetes cluster, there are two types of nodes: the control plane and the worker nodes, with the former being mostly responsible for the administrative functions of the cluster. The latter contains a kubelet agent process responsible for interacting with the control plane. The kubelet ensures that pods are running, enforces resource requirements, and monitors resource usage (e.g. containers' health and performance in terms of CPU, memory, I/O and networking), using cAdvisor, which also makes this data available for scraping for third-party application such as Prometheus. 

Many metrics are exposed by Kubernetes as a whole that might be scrapped by third-party applications, including ones not directly concerning containers (e.g., nodes, storage, etc). The relevant metrics for this work are the ones that concern respectively the usage of memory, the usage of RSS (resident set size), and the memory swap by the container (\texttt{container\_memory\_usage\_bytes},\\\texttt{container\_memory\_rss}, and \texttt{container\_memory\_swap}).

\subsection{Class of Service}

Within a Kubernetes object definition file, CPU and Memory resources may be specified there both in terms of requests and also their allowed usage limits. The former establishes the minimum necessary for executing the object, thus the Kubernetes' scheduler searches for a suitable node for scheduling that has at least the requested resources available. When the pod starts to execute, the limits play a major role by hard enforcing the specified values, therefore the pod can use resources up until the limit values are available. The values of requests and limits play a role in which Quality of Service (QoS) class the pod will be assigned to: \texttt{Guaranteed}, \texttt{Burstable} or \texttt{Best Effort}. In practice, these classes define the priority of resource allocation and execution for the pod. Under pressure, the first tasks to be evicted are the ones in \texttt{BestEffort}, then \texttt{Burstable}, and then \texttt{Guaranteed}. Similarly, \texttt{BestEffort} pods may not use the resources that are available for the Guaranteed pods. 

In cloud-first environments, especially in systems that deal with multi-tenancy, several research papers~ \cite{10.1145/3297858.3304005,10.1145/3581784.3607076} commonly distinguish between two types of workloads: the latency-critical (LC) and the best-effort (BE) ones. The former is often related to user-facing workloads that have a deadline to reply as a slower response time might affect the users' perception of it. Computing workloads, such as high-performance ones, are usually classified as BE because they attempt to make use of all the resources available in the system.



\noindent\hspace{2em}\textbf{Setting limits in the object specification}. In certain environments, multiple workloads often run concurrently on the same virtualized hardware. When all pods are burstable (i.e., they have no CPU or memory limits set) and request the same amount of resources, the Kubernetes scheduler allocates them across nodes up to the maximum capacity (e.g., if a node’s capacity is $x$ and each pod requests $y$, the node can host up to $x/y$ pods). While this may initially appear optimal, it can lead to inefficient resource usage, as actual utilization is often lower than requested, with pods requiring peak resources only during short bursts. By adjusting resource requests to match real usage patterns, the scheduler could accommodate more pods per node. However, without limits, a single misbehaving pod could monopolize system resources, preventing others from bursting as needed and negatively impacting overall application performance. Therefore, imposing resource limits on pods helps prevent such scenarios.

\noindent\hspace{2em}\textbf{Resource Adaptivity.} Kubernetes plays a key role in managing HPC resources in the cloud. Greneche et al. \cite{greneche2022autoscaling} explored HPC workload scaling, while Pupykina and Agosta \cite{pupykina2019survey} surveyed memory management techniques. Autoscalers for cloud workloads are typically proactive or reactive. Proactive scaling anticipates changes in consumption, while reactive scaling responds during or after the event. Many approaches use application-level metrics, such as meeting quality-of-service deadlines, and hardware counters to assess scaling needs. Representative works include Petrucci et al.\cite{petrucci2015octopus} and Chen et al.\cite{10.1145/3297858.3304005}. 


\subsection{Vertical Pod Autoscaler}
The Vertical Pod Autoscaler (VPA) is the default vertical scaling solution in Kubernetes, designed to automate resource configuration and optimize cluster utilization, and dynamically adjust pod resource requests based on usage while maintaining predefined request-limit ratios. By default, VPA is not enabled in standard Kubernetes installations and requires manual activation.

The VPA scales pods by increasing or decreasing resource requests according to historical usage. Its architecture comprises three components: the Recommender, Updater, and Admission Plugin. The Recommender analyzes historical pod usage to model optimal resource allocations. The Updater enforces these recommendations by evicting and restarting non-compliant pods. The Admission Plugin modifies new pod configurations to align with the Recommender's recommendations.

Memory recommendations in VPA maintain a predefined probability threshold for exceeding requests within a specified time window (e.g., 1\% in 24h, per official documentation). Historical execution data is stored for up to 8 days and informs future recommendations. In out-of-memory scenarios, where the requested memory is higher than the available memory, the application restarts with a memory limit equal to its previous request plus a predefined margin (default: 20\%). 


Additionally, updated pods may be rescheduled to different nodes, potentially disrupting multi-node HPC applications and causing failures or undefined states. Furthermore, Figure \ref{fig:normal_vpa} illustrates VPA recommendations for various applications with updating disabled. As it will be discussed later, several workloads exhibit slow VPA adaptation, leading to repeated OOM errors if the recommended updates were enforced. This issue arises because VPA relies on historical patterns, which are inconsistent in HPC workloads due to varying input characteristics. 

Due to frequent OOM errors and potential restarts under standard VPA policies for HPC applications, we propose a model that estimates execution time and total memory footprint for HPC workloads. Further details of this approach are provided in Section \ref{sec:vpa-algo}.

\section{Methodology} \label{sec:meth}
\begin{figure}
     \centering
     \includegraphics[width=\textwidth]{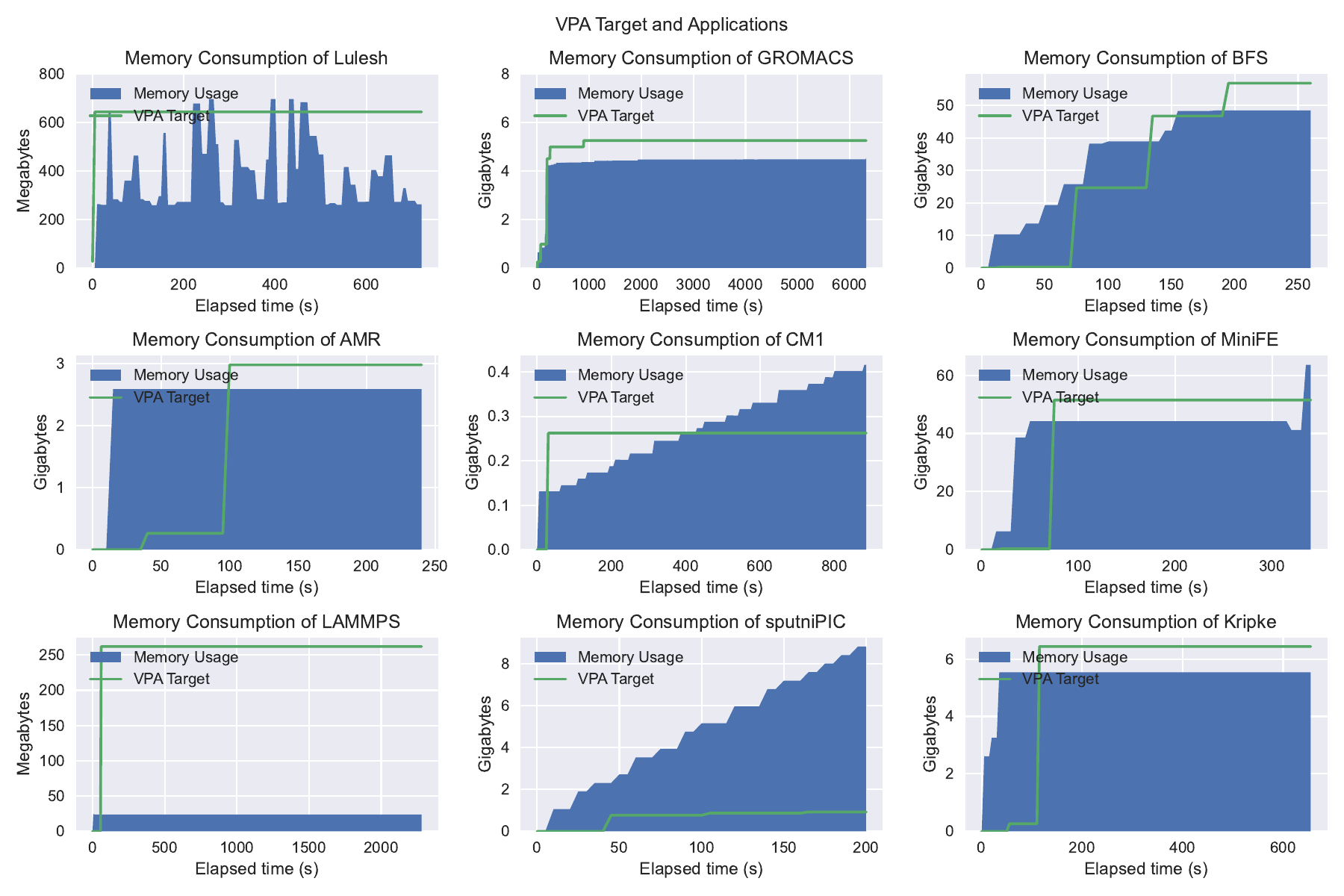}
        \caption{Memory consumption pattern of all applications listed in Section \ref{sec:apps}. The recommendation given by the Vertical Pod Autoscaler is also shown. The data has a sampling time of 5 seconds. }
        \label{fig:normal_vpa}
\end{figure}

In this work, we analyze nine different applications and their respective workloads. These applications are representative of several scientific domains and exhibit distinct consumption patterns, both in terms of quantity and variation. For each listed application, we built a containerized version of them. The typical memory usage patterns of these applications are shown in Figure \ref{fig:normal_vpa}, and a numerical summary of their execution times and memory footprint is presented in Table \ref{tab:numerical-results-raw}.

As the first step of the work, we propose classifying our analyzed applications into two distinct memory consumption patterns: Growth and Dynamic. We define the Growth (G) pattern as a non-decreasing monotonic function, which also serves when the behaviour of memory consumption is stable. However, a caveat of this definition is that, in real measurements, there may exist slightly deviations due to the existing noise in the measurements. Therefore, we still believe that this definition is still valid if the deviation is a value between $\left[ -2\%,+2\% \right]$ of the previous one.

For all other functions that do not satisfy the previous definition, we label them as ``Dynamic" (D) as their behaviour might either follow an apparently periodic pattern or be inherently stochastic, therefore we are not able to establish a general model for them. These functions are normally characterised by having a decrease in memory consumption at some point during their execution time.

\subsection{Applications} \label{sec:apps}
\noindent\hspace{2em}\textbf{AMR}. The Adaptive Mesh Refinement is a technique used in several scientific domains, such as fluid dynamics, to dynamically enhance the solution in certain regions of the simulation. In this work, we used the available MiniAMR proxy application by Mantevo~\cite{heroux2009improving}. We adapted the available two moving spheres problem to 10 OpenMP threads and one MPI rank. 

\noindent\hspace{2em}\textbf{BFS}. Breadth First Search is an algorithm available on the Ligra framework~\cite{shun2013ligra}, a graph processing framework for shared memory, and we use OpenMP for parallel processing on 10 threads. The algorithm itself explores all the vertices in a graph at the current depth before moving to the next one. We use a tool bundled within the Ligra framework (\texttt{rMatGen}) to pre-generate a graph with 100 million vertices that will serve as input that consists of 9.6 GB in terms of file size.

\noindent\hspace{2em}\textbf{CM1}. The Cloud Model 1~\cite{bryan2002benchmark} is a numerical model designed for idealized studies of atmospheric phenomena, such as thunderstorms. We use the default input file that is bundled with the application as well as one MPI rank and 10 OpenMP threads. 

\noindent\hspace{2em}\textbf{GROMACS}. A molecular dynamics application~\cite{abraham2015gromacs,doi:10.1021/acs.jcim.2c00044} that is able to simulate proteins, lipids and nucleic acids.  In this paper, the used benchmark for GROMACS is the ``\texttt{benchRIB}'' provided by the Max Planck Institute for Multidisciplinary Sciences (MPINAT). It consists of a simulation with 2 million atoms (ribosome in water), and the benchmark is executed in both MPI and OpenMP (1 thread per MPI rank, 10 ranks).

\noindent\hspace{2em}\textbf{Kripke}. An application~\cite{kunen2015kripke} designed to be a proxy for an discrete-ordinates transport code, with the focus of studying the performance characteristics of data layouts, programming models, and sweep algorithms. In this work, we use all the default inputs for the application with the exception of the number of groups (640) and the number of iterations (30), 10 OpenMP threads.

\noindent\hspace{2em}\textbf{LAMMPS}. The Large-scale Atomic/Molecular Massively Parallel Simulator is a molecular dynamics code with a focus on materials modeling. There are several example applications that LAMMPS has bundled, and in this work we make use of the HEAT problem that carries simulations of thermal gradients for a Lennard-Jones fluid, with 10 OpenMP threads.   

\noindent\hspace{2em}\textbf{LULESH}. The Livermore Unstructured Lagrangian Explicit Shock Hydrodynamics~\cite{LULESH:spec} is a proxy application designed to study a simple Sedov blast problem with analytic answers. In this work, we use the OpenMP version of LULESH with a size of 90$^3$ as input parameter. Classified in this work as having a dynamic memory consumption pattern, LULESH displays a seemingly chaotic memory consumption pattern including many bursts during short period followed by steep decreases.

\noindent\hspace{2em}\textbf{MiniFE}. The Finite Element proxy application~\cite{heroux2009improving} (MiniFE) is described by its authors as being the "best approximation to an unstructured implicit finite element or finite volume application, but in 8000 lines or fewer". We use the default input with the size (1000, 1000, 1000) in terms of the dimensions (NX, NY, NZ). MiniFE appears to have a growing pattern up until the end of its execution, where there is a steep decrease followed by a steep increase in consumption.

\noindent\hspace{2em}\textbf{sputniPIC}. This is an application used for space plasma simulations that uses the particle-in-cell method~\cite{9235052}. We built a container image for it using MPI, and used the bundled two-dimensional Geospace Environmental Modeling Challenge (GEM2D) as input problem, adapting it for the usage of 10 MPI ranks.

\begin{table}[ht]
\centering
\caption{Features of the application workloads in Section \ref{sec:apps}. Patterns refers to the memory consumption, i.e., Growth (G) or Dynamic (D). The memory footprint was calculated based on the area of the consumption function from Figure \ref{fig:normal_vpa} (in blue).}
\label{tab:numerical-results-raw}
\resizebox{0.5\linewidth}{!}{%
\begin{tabular}{|c|c|c|c|c|}
\hline
\textbf{Application} & \textbf{\begin{tabular}[c]{@{}c@{}}Pattern\end{tabular}} & \textbf{\begin{tabular}[c]{@{}c@{}}Execution \\ Time\end{tabular}} & \textbf{\begin{tabular}[c]{@{}c@{}}Max. \\ Memory\end{tabular}} & \textbf{\begin{tabular}[c]{@{}c@{}}Memory\\ Footprint\end{tabular}} \\ \hline
\rowcolor{lightergray} AMR & G & 253s & 2.6GB & 0.62 TB \\ \hline
BFS & D & 287s & 48.4GB & 9.4 TB \\ \hline
\rowcolor{lightergray} CM1 & G & 913s & 415MB & 0.24 TB \\ \hline
\rowcolor{lightergray} GROMACS & G & 6420s & 4.5GB & 27.18 TB \\ \hline
\rowcolor{lightergray} Kripke & G & 650s & 5.5GB & 3.5 TB \\ \hline
\rowcolor{lightergray} LAMMPS & G & 2321s & 23.7MB & 0.054 TB \\ \hline
LULESH & D & 750s & 696MB & 0.27 TB \\ \hline
MiniFE & D & 352s & 63.7GB & 13.8 TB \\ \hline
\rowcolor{lightergray} sputniPIC & G & 210s & 8.8GB & 1.0 TB \\ \hline
\end{tabular}}
\end{table}

\subsection{In-flight Pod Updates and Swap}
By default, Kubernetes does not allow updating a pod's requests or limits during execution. To modify these settings, the pod must be restarted, which interrupts the workload. However, an alpha feature named ``\texttt{InPlacePodVerticalScaling}'' was recently introduced. By submitting a patch to the pod, the user is able to replace the information regarding the requests and limits of that object. Users may also explicitly specify what should happen when the pod is updated (i.e., not restart the pod or restart it). One of the caveats of this feature is that it is not possible to change the QoS class (as described in Section \ref{sec:background}) of the pod. In practice, this means that a pod in the Best Effort class can be freely resized but will still keep its QoS class as Best Effort even if it satisfies the requirements to be in other classes. 



Another relevant point to mention is that these patches do not necessarily take effect immediately when they are issued. Rather, our empirical evidence suggests that the nominal changes to pod limits are written instantly into the Kubelet; however, there might be a delay of several seconds before these changes synchronize with the actual container and become \textit{de facto} effective. Additionally, we observe that when a patch is issued for a memory limit lower than the current memory usage, the synchronization process is significantly prolonged. Even if swap usage is enabled within Kubernetes, this synchronization stage may not be able to conclude for the entire duration of the application’s execution.

\noindent\hspace{2em}\textbf{Swap}. The usage of swap memory is a new feature in Kubernetes that is leveraged in this work to avoid sudden out-of-memory errors in our applications. The current behaviour of Kubernetes is to fail to start if swap is enabled on the node, hence the feature needs to be manually enabled within Kubernetes and, in the eventual scenario of memory scarcity of a pod, the container will automatically attempt to use the available swap device that is set on the Linux kernel.

Nonetheless, as expected, swap performance strongly depends on the system's storage infrastructure, as this determines the maximum speed for read/write operations (e.g., SSDs are faster than HDDs). A current limitation is the inability to control a per-pod swap limit, meaning the underlying storage infrastructure can easily become bottlenecked if many workloads use swap simultaneously.

\subsection{ARC-V}
\begin{figure}[!t]
    \centering
    \includegraphics[scale=0.4]{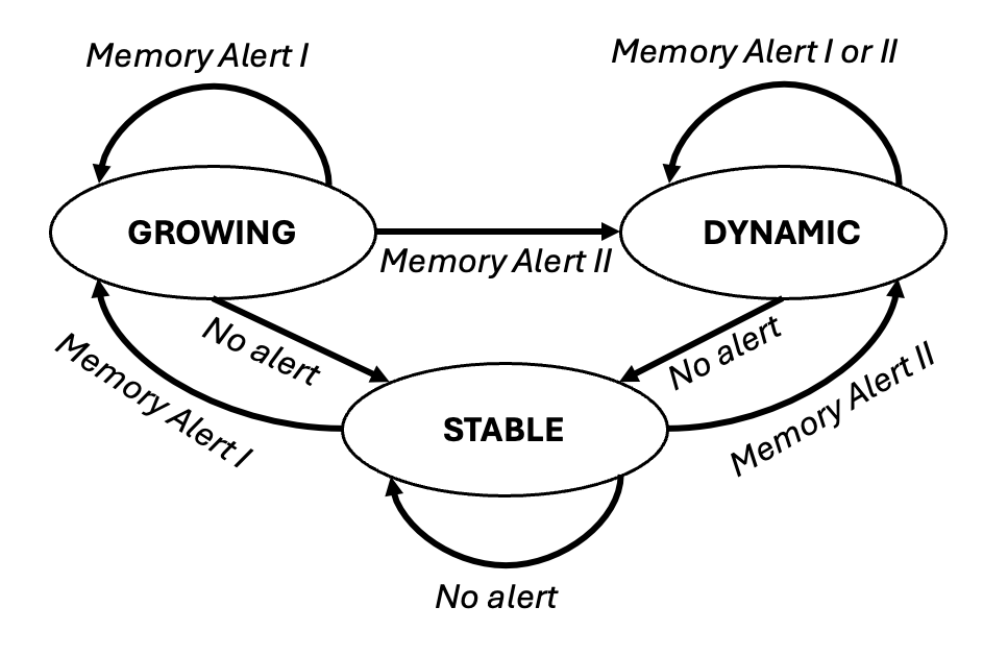}
    \caption{A depiction of the high-level design of the ARC-V autoscaler.}
    \label{fig:arcv-design}
\end{figure}

To establish a vertical autoscaling policy that gives suitable recommendations for HPC workloads on containerized environments, we design the Adaptive Resource Controller - Vertical (ARC-V). This policy, together with its implementation, is established as a state machine that requires specific signals (named ``memory alerts") to move between its different states whereas a depiction may be seen in Figure \ref{fig:arcv-design}. There are two major guiding remarks when designing ARC-V:

\noindent\hspace{2em}\textbf{HPC workloads generally have an initialization phase}. Many applications spend several seconds increasing their memory usage until reaching a degree of stability in consumption (e.g., GROMACS, Kripke, MiniFE). 

\noindent\hspace{2em}\textbf{HPC workloads might have different phases during its execution}. Despite we classify the applications in two broad categories, these categories might also apply through certain stages of the application: an application classified as "Dynamic" might, during certain time, have a "Growth" component in its execution pattern, and vice versa. 

With these guiding remarks in mind, we defined three states in which the application may be located during its execution, primarily because there is no prior knowledge about its consumption pattern. The states were named "Growing", "Dynamic", and "Stable", each with its own scaling policy. Therefore, it is expected that an application will be in one of these states for the majority of its execution.

As its name implies, the "Stable" state means that memory consumption within that particular window remains constant for a long period, or with very little fluctuation, indicating that it is likely possible to reduce memory allocation. Meanwhile, the "Growing" state is defined when the application displays an increase in memory consumption. The "Dynamic" state is directly associated with a decrease in consumption.

The change between these states happens through signals that are sent every predefined timesteps. The collected data is separately analyzed for trends of increase, decrease, or stability in consumption (memory signals I, II, and no signal, respectively). The signal is therefore analyzed by the implementation, and, depending on the current state of the application, different actions are performed. If the application is in the "Growing" or "Stable" state, a single signal is enough to move the application to the "Dynamic" state. However, moving from "Dynamic" to "Stable" requires the absence of signals for an extended period of time, and there is no possible direct transition from the "Dynamic" to "Growing" states..

The scaling policies for each state are defined as follows:
\begin{itemize}
    \item \textbf{Growing}: After a memory signal I, if the difference between the consumption and the actual recommendation is lower than certain threshold, a forecast of the consumption for the next 60 seconds is done and the recommendation is adjusted, else the provided recommendation stays stable. A single signal II moves the application to the "Dynamic" state, and several absence of signals in a row moves the application to the "Stable" state.
    \item \textbf{Dynamic}: As one application enters in this state, there is the need to be very conservative regarding the memory limits as there can be steep spikes. While in this state, the memory decrease is limited to the global maximum that has been achieved by the application,  
     \item \textbf{Stable}: A single signal (either I or II) moves it to either the "Dynamic" or "Growing" states. In case the application persists in the state for several timesteps, the memory allocation decreases in 10\% each time, up to a limit of 102\% of the actual memory usage. This extra 2\% was arbitrarily chosen (Section \ref{sec:impl}) to account for very small variations during the measurements, in which we consider the application still persists in the same state.
\end{itemize}

Finally, we also take swap usage in consideration. Whenever the application makes use of swap due to sudden steep spikes in memory usage in comparison to the current established limit (e.g., the last few seconds of MiniFE), ARC-V also takes into the account the amount of swap used when calculating the new limit, providing enough memory for eventual pages in disk to be transferred to the main memory.

\section{Implementation} \label{sec:impl}
\subsection{VPA Simulator} \label{sec:vpa-algo}
The major comparison metrics utilized in this paper are the ones derived from the Vertical Pod Autoscaler: runtime and the memory footprint. We define the memory footprint of the VPA as the total area below the provided recommendation during certain application execution time (i.e., the green line in Figure \ref{fig:normal_vpa}). However, the initial values of the VPA recommendation policy, as displayed by Figure \ref{fig:normal_vpa}, are not a good parameter for comparison because it uses a bottom-up approach, thus the applications would not even be able to start its execution while the values are lower than the memory usage. Therefore, we established a set of procedures based on the documentation of the VPA to simulate its behaviour. 

\begin{itemize}
    \item The first recommendation of the VPA is zero as it does not have any data regarding of the application. In our algorithm, this is replaced for the first recommendation given instead.
    \item We consider that all recommendations are static and do not change over time unless there is an out-of-memory error in the application. This is the usual pattern seen in our results when the memory usage has been lower than the recommendation.
    \item When the recommendation is lower than the memory usage of the application, we consider that there has been an OOM error issue and the application is restarted with the new recommendation being 20\% higher in comparison to what was requested by the application immediately before it was restarted. This is in agreement with the design documentation and default values of the VPA.
\end{itemize}

\subsection{ARC-V}
The proof-of-concept implementation of our framework is done in Python using the Kubernetes Python API, with the data being retrieved directly from an Kubernetes end point.

\noindent\hspace{2em}\textbf{Parameters}. There are several parameters that were defined and would potentially affect the execution. In particular, we define the "stability factor" as 2\%, meaning that the window of collected values may fluctuate between [-2\%, 2\%]. The change in the value may affect for how long one application might be considered within the "Stable" state, therefore affecting the overall increase/decrease of the provided recommendation from ARC-V. The number of collected metrics as part of the measurement window is also a factor. Finally, as it might take some time for the changes in memory limits to be enforced, we set up a timeout of 60 seconds before issuing a new decision regarding the changes of states.

\noindent\hspace{2em}\textbf{Signals}. Earlier implementations of the signals relied on linear regressions to see whether there is any trend in the window of measurements, however empirical results found this technique not reliable when dealing with small windows of data and/or abrupt changes of values. Our current implementation relies on the sorting of the elements: a non-sorted order (signal II) means that there might have been a decrease in the window, while a sorted order may possibly mean an increase (signal I) or stability (if all elements are equal, no signal). However, the forecasting for Growing mode, when the difference between the memory consumption and the recommendation policy is small, is performed through linear regression.

\noindent\hspace{2em}\textbf{Initialization assumption and automatic classification}. For the implementation, we assume that the pod has more than enough memory to execute through the initialization phase, which is parametrized to 60 seconds. This time is necessary for ARC-V to start analyzing the application and defining which state it should place it, based on how the collect data for consumption behaves. Furthermore, while the application will not suffer from out-of-memory error, as swap memory usage is enabled in the cluster, it is very likely that the application would suffer a strong performance degradation through its execution from the beginning, thus the requirement for memory. For the experiments in this work, the initial memory requests/limits were set as 20\% of the maximum used in the node (i.e., Table \ref{tab:numerical-results-raw} for reference) but could be set to any value as long as the condition discussed above is satisfied. ARC-V would eventually adjust if set too high, and decrease if too low.

\section{Evaluation} \label{sec:eval}
\begin{figure}[ht]
     \centering
     \includegraphics[width=\textwidth]{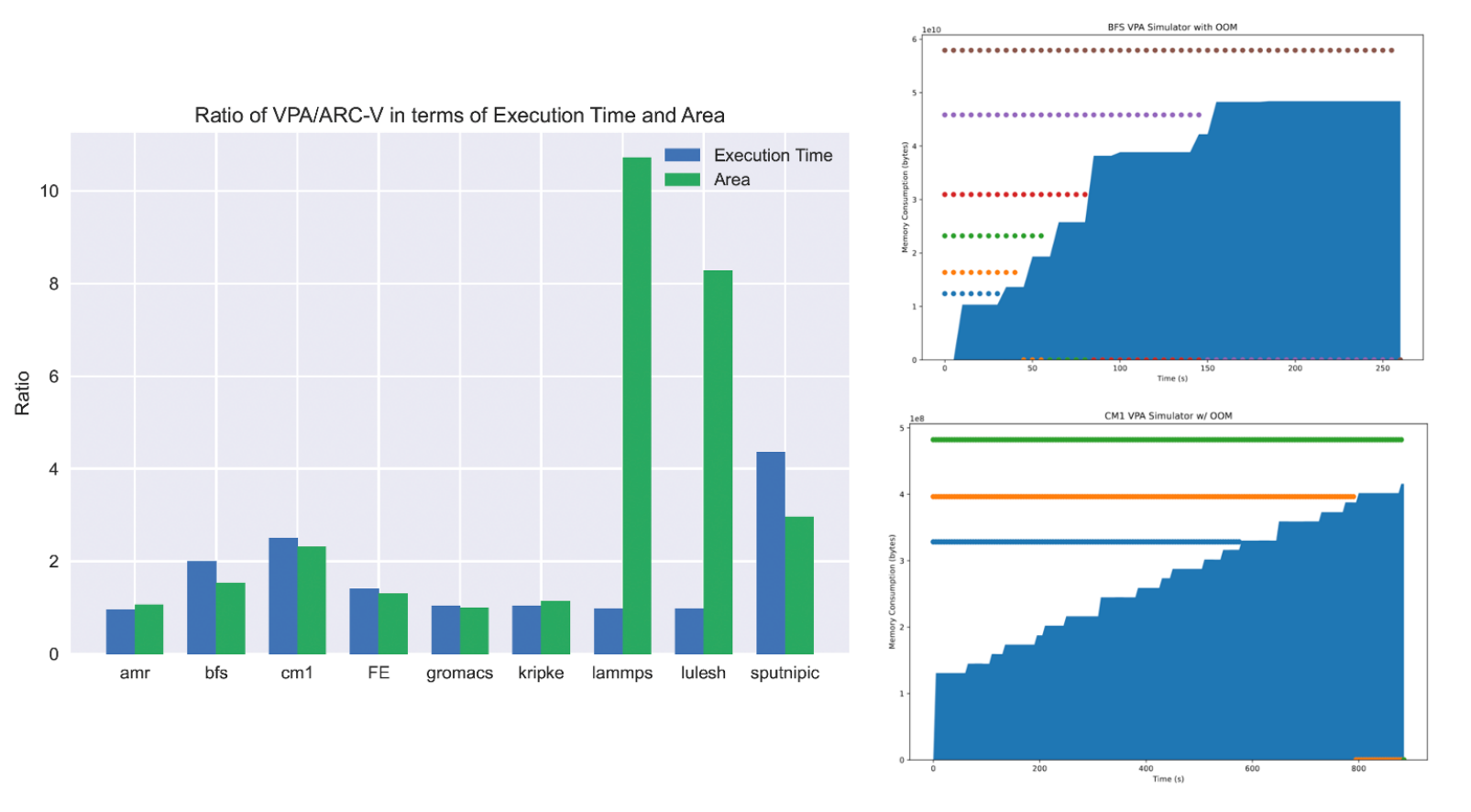}
        \caption{On left, the ratios between the values of memory footprint and execution time of VPA and the ARC-V policy. On right, a depiction of how the VPA simulator works: every time the recommendation is lower than the actual usage, the application needs to restart with 20\% more memory. }
        \label{fig:arcv_vs_vpa}
\end{figure}
\noindent\hspace{2em}\textbf{Infrastructure}. We use three nodes from the CloudLab infrastructure \cite{Duplyakin+:ATC19} for running our experiments. A single node consists of a dual Intel E5-2660 10-core CPUs, 256 GB DDR4 RAM, 2x 1TB mechanical disks running at 7200 RPM. Each node executes Ubuntu 22.04, while Kubernetes, with swap usage enabled, is installed within all nodes using the K3s distribution v1.29.6, with one control plane and two working nodes, where VPA v1.29.3 is deployed. 

\noindent\hspace{2em}\textbf{ARC-V vs VPA}.
Figure~\ref{fig:arcv_vs_vpa} displays all the results obtained with ARC-V compared to the simulated VPA policy that is established for Kubernetes and cloud applications. For MiniFE, that uses swap memory just at the end of its execution, we do not count the swap as provisioned memory in ARC-V since the resource being used in this case is disk. 

\begin{figure*}
     \centering
     \includegraphics[width=\textwidth]{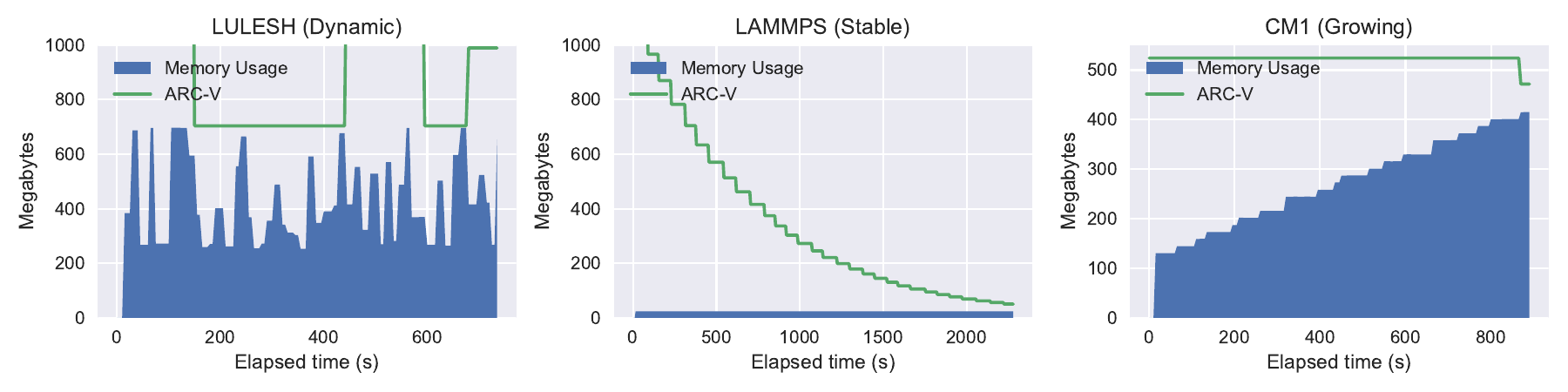}
        \caption{Example cases of ARC-V defining the memory limits for applications that are dominated mostly by a certain pattern. In the LULESH and LAMMPS cases, the starting values are higher than the actual consumption and not displayed in the plot for scaling purposes. The starting values in this plot are exaggerated (in comparison to the experiments done in Figure~\ref{fig:arcv_vs_vpa}) for display purposes.}
        \label{fig:interesting2}

\end{figure*}

\noindent\hspace{2em}\textbf{Overhead}. Our implementation executes in another node than the one that executes the container. It runs on user space level, requiring only Kubernetes access permissions, and is not containerized. Furthermore, the time steps for scrapping the data from the running node's kubelet is large (5 seconds), decreasing the chance of possible performance impact in any shared resources. With the exception of MiniFE (which uses swap in order to avoid OOM during certain phases), we observe that the difference of the execution times are usually below 3\%.

\noindent\hspace{2em}\textbf{Memory provisioning}. In all cases, there are major resource savings in terms of consumption, with they being expressive in applications that spend most of the time in "Growing" state, such as CM1 and sputniPIC. We display three example decisions of ARC-V in Figure~\ref{fig:interesting2} for applications that are dominated mostly by a certain state. 

In the particular case of LAMMPS, the difference of over 10 times is because the VPA allocates automatically too much memory and do not resize its recommendation through the execution time of the application,  while ARC-V quickly detects the stable pattern and starts to get closer to the actual usage. The great difference is mostly because LAMMPS makes a small usage of memory. In contrast, AMR displays a stable pattern as well but the ratio between VPA/ARC-V is about 1.06 as AMR makes a large use of memory.

\noindent\hspace{2em}\textbf{Execution time}. The  enforcement of the VPA policy under the assumption of no checkpointing leads to a major increase in its execution time as the application needs to restart several times due to OOM errors as the VPA recommendation falls below the requested memory by the application (i.e., the right part of Figure~\ref{fig:arcv_vs_vpa}), while ARC-V avoids OOM by using a top-down approach, and in the case of MiniFE, swap. This effect is strongly seen in applications that are dominated by the Growing state, such as BFS, CM1, and sputniPIC.  


\noindent\hspace{2em}\textbf{Use cases}. The memory savings are important for certain applications, such as Kripke. The recommended value quickly decreases from 6.6GB  (20\% of maximum memory, initial request) to 5.6GB (around 16\%) at around 1/3 of its execution. Such savings would enable one to run concurrently other applications with the same workloads in this work such as CM1, Lulesh or LAMMPS. However, discussing potential effects of resource sharing is out of scope of this paper. 

\section{Conclusions}
In this work, we have identified that the current VPA is inadequate for HPC workloads that exhibit dynamic memory usage, potentially generating resource waste and out-of-memory errors. To address these limitations, we proposed \tool to improve elastic memory scaling in HPC applications when running on cloud environments. \tool addresses bursty memory usage in HPC workloads by auto-classifying the memory consumption in three different modes, monitors performance metrics, and proposes a recommendation policy in a top-down approach, while also leveraging swap memory accordingly at runtime when necessary. We evaluated \tool in nine HPC applications and the results show that \tool can effectively reduce memory waste and improve performance compared to the standard Kubernetes VPA. 

\section{Acknowledgements}
This research is supported by the European Commission under the Horizon project OpenCUBE (GA-101092984) and SESSI, SeRC Efficient Simulation Software Initiative for HPC Malleability.
\bibliographystyle{IEEEtran}
\bibliography{main}

\begin{thebibliography}{10}
\providecommand{\url}[1]{#1}
\csname url@samestyle\endcsname
\providecommand{\newblock}{\relax}
\providecommand{\bibinfo}[2]{#2}
\providecommand{\BIBentrySTDinterwordspacing}{\spaceskip=0pt\relax}
\providecommand{\BIBentryALTinterwordstretchfactor}{4}
\providecommand{\BIBentryALTinterwordspacing}{\spaceskip=\fontdimen2\font plus
\BIBentryALTinterwordstretchfactor\fontdimen3\font minus \fontdimen4\font\relax}
\providecommand{\BIBforeignlanguage}[2]{{%
\expandafter\ifx\csname l@#1\endcsname\relax
\typeout{** WARNING: IEEEtran.bst: No hyphenation pattern has been}%
\typeout{** loaded for the language `#1'. Using the pattern for}%
\typeout{** the default language instead.}%
\else
\language=\csname l@#1\endcsname
\fi
#2}}
\providecommand{\BIBdecl}{\relax}
\BIBdecl

\bibitem{peng2021holistic}
I.~Peng \emph{et~al.}, ``A holistic view of memory utilization on hpc systems: Current and future trends,'' in \emph{Proceedings of the International Symposium on Memory Systems}, 2021, pp. 1--11.

\bibitem{egwutuoha2012proactive}
I.~P. Egwutuoha \emph{et~al.}, ``A proactive fault tolerance approach to high performance computing (hpc) in the cloud,'' in \emph{2012 Second International Conference on Cloud and Green Computing}.\hskip 1em plus 0.5em minus 0.4em\relax IEEE, 2012, pp. 268--273.

\bibitem{jin2012checkpointing}
H.~Jin \emph{et~al.}, ``Checkpointing orchestration: Toward a scalable hpc fault-tolerant environment,'' in \emph{2012 12th IEEE/ACM International Symposium on Cluster, Cloud and Grid Computing (ccgrid 2012)}.\hskip 1em plus 0.5em minus 0.4em\relax IEEE, 2012, pp. 276--283.

\bibitem{10.1145/3297858.3304005}
S.~Chen \emph{et~al.}, ``Parties: Qos-aware resource partitioning for multiple interactive services,'' in \emph{Proceedings of the 24th International Conference on Architectural Support for Programming Languages and Operating Systems}, ser. ASPLOS '19.\hskip 1em plus 0.5em minus 0.4em\relax ACM.

\bibitem{10.1145/3581784.3607076}
B.~Aksar \emph{et~al.}, ``Prodigy: Towards unsupervised anomaly detection in production hpc systems,'' in \emph{Proceedings of the International Conference for High Performance Computing, Networking, Storage and Analysis}, ser. SC '23.\hskip 1em plus 0.5em minus 0.4em\relax ACM, 2023.

\bibitem{greneche2022autoscaling}
N.~Greneche and C.~Cerin, ``Autoscaling of containerized hpc clusters in the cloud,'' in \emph{2022 IEEE/ACM International Workshop on Interoperability of Supercomputing and Cloud Technologies (SuperCompCloud)}.\hskip 1em plus 0.5em minus 0.4em\relax IEEE, 2022, pp. 1--7.

\bibitem{pupykina2019survey}
A.~Pupykina and G.~Agosta, ``Survey of memory management techniques for hpc and cloud computing,'' \emph{IEEE Access}, vol.~7, pp. 167\,351--167\,373, 2019.

\bibitem{petrucci2015octopus}
V.~Petrucci \emph{et~al.}, ``Octopus-man: Qos-driven task management for heterogeneous multicores in warehouse-scale computers,'' in \emph{2015 IEEE 21st International Symposium on High Performance Computer Architecture (HPCA)}.\hskip 1em plus 0.5em minus 0.4em\relax IEEE, 2015.

\bibitem{heroux2009improving}
M.~A. Heroux, D.~W. Doerfler \emph{et~al.}, ``{Improving Performance via Mini-applications},'' Sandia National Laboratories, Tech. Rep. SAND2009-5574, 2009.

\bibitem{shun2013ligra}
J.~Shun and G.~E. Blelloch, ``Ligra: a lightweight graph processing framework for shared memory,'' in \emph{Proceedings of the 18th ACM SIGPLAN symposium on Principles and practice of parallel programming}, 2013, pp. 135--146.

\bibitem{bryan2002benchmark}
G.~H. Bryan and J.~M. Fritsch, ``A benchmark simulation for moist nonhydrostatic numerical models,'' \emph{Monthly Weather Review}, vol. 130, no.~12, pp. 2917--2928, 2002.

\bibitem{abraham2015gromacs}
M.~J. Abraham \emph{et~al.}, ``{GROMACS: High performance molecular simulations through multi-level parallelism from laptops to supercomputers},'' \emph{SoftwareX}, vol.~1, pp. 19--25, 2015.

\bibitem{doi:10.1021/acs.jcim.2c00044}
C.~Kutzner \emph{et~al.}, ``{GROMACS in the Cloud: A Global Supercomputer to Speed Up Alchemical Drug Design},'' \emph{Journal of Chemical Information and Modeling}, vol.~62, no.~7, pp. 1691--1711, 2022, pMID: 35353508.

\bibitem{kunen2015kripke}
A.~J. Kunen \emph{et~al.}, ``Kripke-a massively parallel transport mini-app,'' Lawrence Livermore National Lab.(LLNL), Livermore, CA (United States), Tech. Rep., 2015.

\bibitem{LULESH:spec}
``{H}ydrodynamics {C}hallenge {P}roblem, {L}awrence {L}ivermore {N}ational {L}aboratory,'' Tech. Rep. LLNL-TR-490254.

\bibitem{9235052}
S.~W.~D. {Chien} \emph{et~al.}, ``{sputniPIC}: An implicit {Particle-in-Cell} code for multi-{GPU} systems,'' in \emph{2020 IEEE 32nd International Symposium on Computer Architecture and High Performance Computing (SBAC-PAD)}, 2020, pp. 149--156.

\bibitem{Duplyakin+:ATC19}
D.~Duplyakin \emph{et~al.}, ``The design and operation of {CloudLab},'' in \emph{Proceedings of the {USENIX} Annual Technical Conference (ATC)}, Jul. 2019, pp. 1--14.

\end{thebibliography}

\end{document}